\documentclass[letterpaper]{article}

\usepackage[T1]{fontenc}

\usepackage{geometry}
\usepackage{setspace}

\usepackage[style = chem-acs, articletitle=true, doi=true]{biblatex}
\usepackage{graphicx}
\usepackage{float}
\newfloat{scheme}{htbp}{los}
\floatname{scheme}{Scheme}
\floatname{chart}{Chart}
\newfloat{graph}{htbp}{loh}
\usepackage{amsmath}
\usepackage{upgreek}
\usepackage{hyperref}
\usepackage{booktabs}
\usepackage{tabularx}

\usepackage{array}
\usepackage{ragged2e}

\newcolumntype{P}[1]{>{\RaggedRight\arraybackslash}p{#1}}
\newcolumntype{Y}[1]{%
  >{\hsize=#1\hsize\linewidth=\hsize
    \RaggedRight\arraybackslash}X}
\usepackage{chemformula} % Formulas using \ch{}
\usepackage[version = 4]{mhchem} % Formulas using \ce{}

\usepackage{authblk}
\author[1,2]{Sam T.P. Borman*}
\author[2]{Daphne M. Dekker}
\author[2]{Evelijn Akerboom}
\author[2]{Albert Polman}
\affil[1]{Advanced Research Center for
Nanolithography, 1098 XG Amsterdam, The Netherlands}
\affil[2]{NWO-institute AMOLF, 1098 XG, Amsterdam, The Netherlands}

\title{A library-based approach for inverse reconstruction of cavity depth using angle-resolved cathodoluminescence}
\date{*Email: S.Borman@amolf.nl}

\begin{document}

\maketitle

\begin{abstract}
In nanofabrication processes, accurate metrology is a crucial step for obtaining control over the required critical dimensions in a device. Traditionally, scanning electron microscopy (SEM) is used to obtain high resolution measurements of the lateral dimension, but obtaining depth information on a thick sample is more complex. In this work, we use angle-resolved cathodoluminescence (AR-CL) spectroscopy to resolve this third dimension. Using 30 keV electrons in the SEM, CL is created at the bottom of a micrometer scale cavity milled in a gold surface. We show that reflections of the CL off the cavity sidewalls produce an interferogram which is a sensitive probe for the cavity geometry. The interferograms can be modelled using the finite-difference time-domain method, and we demonstrate that a neural network trained on a library of these simulated patterns can be used to reconstruct the cavity depth. This approach establishes a proof of concept for AR-CL based depth reconstruction within a prescribed geometry and could form a route towards advanced three-dimensional metrology of complex nanostructures. 

\end{abstract}

\begin{figure}[H]
    \centering
    \includegraphics[width=0.7\textwidth]{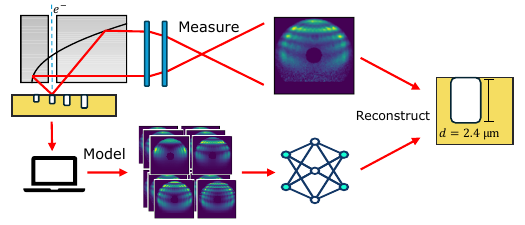}
\end{figure}

\newpage
\section{Introduction}
Advances in semiconductor technology, integrated photonics and photovoltaics generate a need for advanced characterization techniques. The materials and geometry of such devices play a crucial role in their speed, reliability and power consumption \cite{orji_roadmap}, warranting the research of methods to probe their structure and dimensions. Both optical and electron microscopy are well-established methods for dimensional metrology. Scatterometry is widely used to reconstruct grating geometry \cite{bodermann_scatterometry} and confocal profilometry can be used for mapping surface topography, roughness and step edges \cite{kim2021_confocal}. Complementary to these optical techniques, electron-beam methods provide localized information both on device geometry and material composition.

Scanning electron microscopy (SEM) uses a beam of high energy (1-30 keV) electrons to create a highly localized excitation point where secondary electrons (SEs) are collected to create a topographic contrast map of the features\cite{SEM_1983}. The sub-nm resolution of the SEM gives precise information on lateral dimensions, but to obtain information on the depth of a feature is more complex\cite{orji_roadmap, degoede_medsem}. In industrial applications a highly-specialized critical dimension (CD) SEM employs a tilted electron beam (e-beam) to measure height and sidewall angle through the parallax principle \cite{valade_tilted_beam, zhang_finfet_tilt}. In more conventional SEMs the sample itself can be tilted and analyzed using a goniometric construction but for larger samples this is not always possible and it can prove challenging to maintain the strict beam alignment while tilting the stage. Full three-dimensional (3D) reconstruction of nanostructures can be realized with transmission electron microscope (TEM) tomography \cite{midgley2009}, but this is a highly specialized method requiring careful preparation of thin lamella samples. Another metrology method which is gaining more relevance as features get smaller is atomic force microscopy (AFM), which uses a mechanical probe following the surface topography to quantitatively map features with sub-nm resolution \cite{afm_spie, franz_afm}. Compared to these methods however, SEM measurements are faster, easy to use and more scalable, which motivates development of 3D metrology on this platform. 

In the SEM, the electron beam not only generates SEs, but also light (cathodoluminescence, CL). The CL spectrum is characteristic for the excited material and thus offers a powerful probe of the material's composition, defects and structural features \cite{ebel_CLatlas}. The e-beam thus serves as a highly localized source of light that may provide information complementary to the conventional topographical SE image. This motivates investigating whether CL can provide additional information about structural depth of a feature without tilting the beam or the sample.

Previous work has shown 3D reconstructions of plasmonic near fields that were obtained by taking a tomographic sample tilt series using CL \cite{dionne2015} and electron energy-loss spectroscopy (EELS) \cite{midgley2013}. These methods, however, do not measure geometry but the optical near fields, and they require sample tilting which, as described above, can be impractical. Other studies have demonstrated angle-resolved (AR) CL \cite{coenen_arcl, coenen_arcl2} interferometry as a method for obtaining quantitative geometry information. In \cite{schilder_holography} Schilder et al. retrieve the phase distribution and the lateral distance between the electron beam and a nanoscatterer from the interference between transition radiation and scattered surface plasmon polaritons (SPPs). Akerboom et al. \cite{akerboom_interferometry} extend this to measurements in the axial dimension by placing a plasmonic pillar in the electron beam path. From the interference between the tip and the substrate the time difference between the two excitations is extracted which, using the known electron speed, yields a measure for the tip height. Related work by Taleb et al. \cite{taleb2023_interferometry} demonstrated spectral CL interference with electron-generated light and localized electron excitations, enabling retrieval of relative phase and amplitude of the sample radiation. These studies show how CL interferometry can provide geometric and optical information, but they rely on the coherent excitation of two CL sources and require careful placement of plasmonic scatterers or excitation structures.

In this work we investigate whether it is possible to retrieve axial dimensions, using AR-CL, of a multi-scattering geometry. As a model system, we use a square cavity carved into a monocrystalline gold surface. We limit ourselves to relatively large cavities in the micrometer range for which we record interferences in the transition radiation (TR) that is created as 30 keV electrons excite the gold surface at the bottom of the cavity \cite{javier_bible, brenny2016}. We find that the interferences originate from radiation induced by the direct e-beam and its subsequent reflection from the cavity sidewalls. The fringe density encodes lateral separation between the excitation position and the sidewall, while fringe curvature and shadowing effects carry information on the cavity depth. The depth dependence of these features is subtle but we find that it can be inferred from the interferograms using inverse analysis with correct prior assumptions on the geometry. 

Library-based machine-learning approaches have previously been used to retrieve two-dimensional light source positions from experimentally measured AR-CL radiation patterns \cite{buijs_library}. Here, we extend from lateral source localization to depth reconstruction within a prescribed cavity geometry by taking a complementary approach: using the finite-difference time-domain (FDTD) simulation method we construct a library of simulated interferograms for varying excitation position and cavity depth. By training a convolutional neural network on this library we infer the depth of the cavity with a root-mean-square error (RMSE) of $\sim130$ nm relative to independently obtained tilted-SEM and AFM reference measurements. Although absolute reconstruction accuracy could not be demonstrated due to uncertainty in the reference depths, this work provides a proof of concept that quantitative information on depth of a 3D geometry can be measured using AR-CL without tilting. Future studies could focus on extending the inferred geometry with additional parameters like sidewall angle or surface roughness and investigating whether they can be distinguished from the depth.

\section{Results and discussion}

\begin{figure}[h]
    \centering
    \includegraphics[width=3.33in]{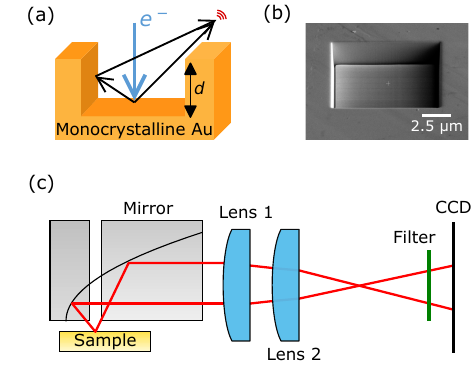}
    \caption{Sample geometry and AR-CL setup. (a) Schematic of the experiment. A 30 keV electron creates CL at the bottom of a cavity of unknown depth, $d$, which scatters out of the cavity and interferes in the far field. (b) SEM image of a $10\times10$ $\upmu$m cavity imaged under a 52$^\circ$ tilt angle. A measurement of the projected sidewall dimensions in the image yields a depth estimate of $\sim2.4$ $\upmu$m. (c) Schematic of the AR-CL measurement setup. The sample is positioned in the focal point of the parabolic mirror and the emitted light is collimated and projected onto a CCD camera through a $550\pm25$ nm optical filter. Using the known geometric properties of the mirror each pixel can be mapped onto a unique set of emission angles $\phi$ and $\theta$.}
    \label{fig1}
\end{figure}

\autoref{fig1}a shows the used sample geometry, where the electron beam excites transition radiation at the bottom of a cavity which scatters from the sidewalls, with the different scattering components interfering in the far field. Cavities of $10\times10$ $\upmu$m width and varying depth were fabricated through focused ion beam (FIB) milling, using 30 keV Ga ions with a total dose of 3.56 nC/$\upmu$m$^2$. One such cavity, imaged under a 52$^\circ$ tilt angle, is shown in \autoref{fig1}b. In this case, the reference depth could be estimated to be approximately 2.4 $\upmu$m by tilting the sample 52$^\circ$ and measuring the projected sidewall. The setup used for CL collection is shown in \autoref{fig1}c. In AR measurements, every pixel on the CCD camera has a unique relation to the polar ($\phi$) and azimuthal ($\theta$) emission angle from the sample. An optical filter is used to collect 550 nm light over a bandwidth of 50 nm. 

\begin{figure}[h]
    \centering
    \includegraphics[width=6.5in]{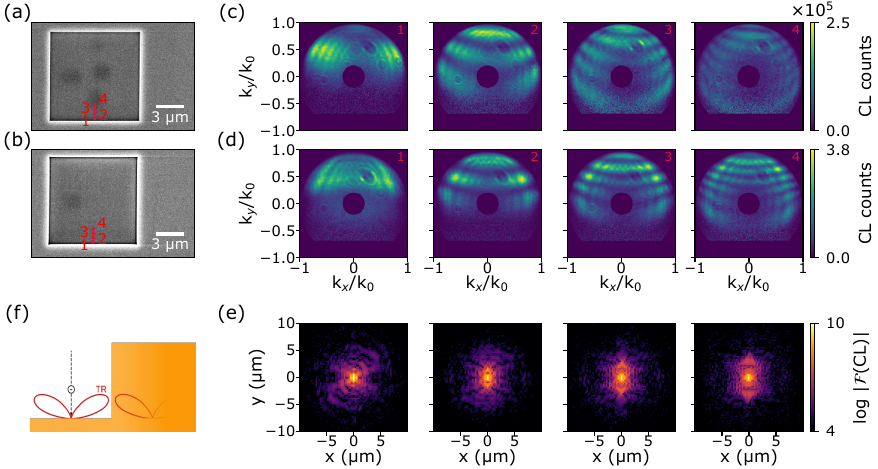}
    \caption{CL interferometry data for two $10\times10$ $\upmu$m cavities with different depth. (a-b) SEM images of cavities of 0.9 $\upmu$m and 2.4 $\upmu$m deep. When imaging without tilt this depth difference cannot be inferred from the SE images alone. Red markers indicate the e-beam impact position where AR-CL data was collected. (c-d) AR-CL interferograms collected at the marked positions in cavities a and b, respectively. The e-beam position impacts the fringe density while the depth difference impacts fringe curvature and shadowing effects. (e) Fourier transform of (d) from reciprocal space to real space. The position of the off-center maximum approximately measures twice the separation between the e-beam position and the lower sidewall. (f) Schematic of transition radiation and its mirror image in the cavity sidewall. The origin of the interference fringes is explained by the primary TR interfering with its reflection.}
    \label{fig2}
\end{figure}

To investigate the relation between the AR-CL interferogram and cavity geometry, two cavities with depths of 0.9 and 2.4 $\upmu$m were investigated. \autoref{fig2}a shows the top view of the 0.9 $\upmu$m cavity taken with SE imaging. The AR-CL images were recorded by placing the electron beam at four different positions on the vertical symmetry axis inside the cavity, for each of which the AR-CL image is shown in \autoref{fig2}c. Several significant trends can be observed from the interferograms. For the electron beam placed closest to the bottom sidewall, an angular emission maximum is observed at $k_y>0$, superimposed on finer interference features along the $k_x$-direction. For $k_y<0$ no interference is observed due to a shadowing effect that blocks the direct TR emission. As the electron beam is moved further inward, a larger number of fringes in the vertical direction appears. The frequency of the fine features along $k_x$ does not appear to change. Also the shadowing effect disappears as the electron beam is moved further away from the edge.

\autoref{fig2}(b,d) show the results for the same experiment on a 2.4 $\upmu$m deep cavity. Several trends are observed compared to the shallower cavity. First of all, the curvature of the fringes is smaller while the number of fringes remains the same. Second, we observe a more extensive shadowing effect for this deeper cavity. These data indicate that fine features in the AR-CL patterns carry information on the cavity geometry.

To analyze the origin of the different interference bands, we take the Fourier transform of the angular patterns in \autoref{fig2}d, which are shown in \autoref{fig2}e. Aside from the central peak, we observe a second-order maximum that shifts away from the center for increasing separation between the electron beam and the sidewall. The characteristic distance can be derived from these Fourier transforms and approximately measures twice this separation, indicating that the origin of the interference fringes is due to the primary generated transition radiation and its reflection from the gold sidewall, as illustrated in \autoref{fig2}f  (see section 1 of the Supporting Information). This analysis is enabled by the broad angular distribution of transition radiation which is composed of two upward tilted angular lobes that in our geometry interact strongly with the sidewalls \cite{javier_bible}. The shadowing effect is then attributed to the fact that a significant fraction of this angular range is not collected by the parabolic mirror, which decreases with increasing distance between the excitation point and the sidewall, and increases with depth as seen when comparing \autoref{fig2}c with \autoref{fig2}d. 

The fact that both the fringe curvature and shadowing effects depend on the cavity depth indicates that these AR-CL measurements contain information that could be used to determine the cavity depth. However, extracting this information directly from the interferograms is challenging, as several scattering and interference effects contribute to the angular distribution. We therefore use an inverse analysis approach, in which a convolutional neural network is trained on a library of simulated interferograms. A forward model is constructed by simulating the emission of transition radiation as a $z-$oriented point dipole source at the bottom of the cavity \cite{kuttge2009} using the finite-difference time-domain (FDTD) method. The angular emission profile is then recorded using a near-to-far field conversion $k$-space monitor. By limiting the excitation position in experiment and simulation to the vertical symmetry axis, we only have to simulate the electromagnetic fields in one half of the cavity and apply mirror symmetry to find the fields for the full cavity, reducing simulation cost. In order to account for the 50 nm bandwidth optical filter used in the experiment the angular emission profile is simulated between 525 and 575 nm and averaged. This partially washes out the interference fringes and reduces their visibility. Since convolutional neural networks can extract information from fine features like edges and curvature, accounting for this contrast reduction in the forward model is important to prevent a mismatch between simulated and experimental interferograms that could reduce the accuracy or precision of depth reconstruction.

We then compose a library of 512 simulated angular scattering patterns, varying the depth of the cavity in the range of 0.5 - 3.5 $\upmu$m and the distance between excitation position and cavity sidewall from 0.2 to 4.0 $\upmu$m. The remaining geometrical parameters are kept fixed, assuming an ideal $10\times 10$ $\upmu$m square cavity with vertical sidewalls. This reduces the dimensionality of the inverse problem, however it also restricts the reconstruction to cavities that fall within the assumed geometry.

\begin{figure}[H]
    \centering
    \includegraphics[width=6.5in]{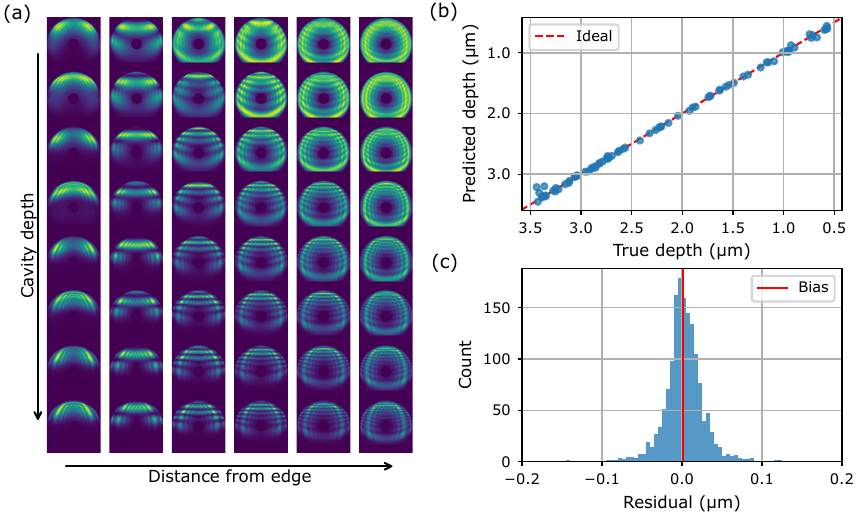}
    \caption{Neural network training and depth inference on a library of simulated AR-CL interferograms. (a) Subset containing 48 of a total of 512 simulated images with varying cavity depth and separation between the e-beam and cavity sidewall. The simulations confirm how fringe curvature and shadowing effects vary with depth while fringe density varies with separation. (b) Depth inference on a set of test images using a single trained network (random seed 19). The strong correlation indicates the ability of the network to interpolate within the parameter space of the training data. (c) Histogram of the test residuals pooled over 20 independently trained networks, each initialized with a different random seed. The red line is the mean of the residuals and indicates the bias of the network analysis.}
    \label{fig3}
\end{figure}

A selection of these simulations is shown in \autoref{fig3}a. Over this wide range of parameters the simulation library shows the same trends as found in the experimental results above; the fringe density changes with the e-beam position and the fringe curvature and shadowing effects change with cavity depth. To infer cavity depth from an AR-CL image, a convolutional neural network is trained on the simulation library. All images are normalized to their maximum intensity and the library is randomly split into 70\% training, 15\% validation and 15\% test data. We train a ResNet-18 neural network \cite{resnet2016} on the training data and track its convergence over a course of 50 epochs using the validation data (see section 2 of the Supporting Information). Out of the 50 epochs, the model with the lowest validation loss is then applied to the test images, which have been kept out of the training loop. The comparison between depth predictions and the ground truth input of the simulations for one trained network is plotted in \autoref{fig3}b. The strong correlation between predicted and true depth demonstrates that the network successfully predicts cavity depth for previously unseen AR-CL images within the simulated parameter space. To assess the reconstruction performance, we repeat the training using 20 different random seeds to initialize the training setup. For each inference result on the test set we compute the residuals, defined as $d_\mathrm{true}-\hat d_\mathrm{pred}$. The histogram of the pooled residuals from all network initializations is shown in \autoref{fig3}c. From these pooled residuals we find a bias of $b = 0.83$ nm, defined as the mean of the residuals, and a residual spread of $\sigma = 37.37$ nm. Combining bias and residual spread yields a root-mean-square-error ($\mathrm{RMSE}=\sqrt{b^2+\sigma^2}$) of 37.38 nm, quantifying the overall reconstruction performance of this network analysis. 

\begin{figure}[H]
    \centering
    \includegraphics[width=2.6in]{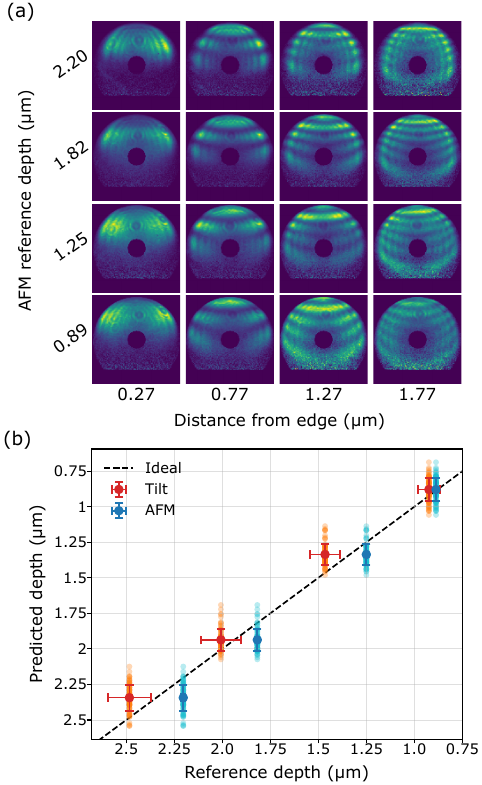}
    \caption{Depth inference on experimental AR-CL data. (a) Normalized AR-CL images, taken at 4 different positions inside 4 different cavities of varying depth. The reference depth was estimated from AFM measurements. (b) Results for depth inference on the experimental AR-CL images, compared to tilted-SEM and AFM as independent references. Orange (tilted-SEM) and cyan (AFM) data points show 320 individual predictions for the 16 measurements evaluated using 20 independently trained networks, while red and blue data points show their average. Vertical error bars represent the standard deviation per cavity, including both excitation position and training-seed variation. Horizontal error bars represent the half-range of manual min/max estimates for tilted SEM or the variation in AFM height profiles. The dashed line indicates perfect agreement between predicted and reference depths.}
    \label{fig4}
\end{figure}

Next, we test the simulation-trained model on an experimentally measured dataset composed of 16 AR-CL measurements taken on 4 different cavities with depths between 0.9 and 2.5 $\upmu$m, each for 4 different e-beam positions on the vertical symmetry axis. Because the library was trained on normalized data, these images are normalized to their maximum intensity as well. The normalized interferograms are shown in \autoref{fig4}a. The reference depths for these cavities were estimated using both SE images of the sidewall projection taken at a 52$^\circ$ tilt angle and atomic force microscopy (AFM) measurements (see section 3 of the Supporting Information). The same 20 independently trained networks are then evaluated on all 16 experimental interferograms, yielding a total of 320 depth predictions. These individual predictions, and their mean value per cavity, are plotted against both the tilted-SEM and the AFM reference depths in \autoref{fig4}b. The reconstructed depths follow the trend of the reference measurement, providing an encouraging proof of concept for AR-CL based depth reconstruction within the prescribed geometry.

The depth reconstruction results are summarized in \autoref{tab:depth_statistics}, showing the mean predicted depth and its standard deviation for each cavity, together with the pooled residual statistics. The reported cavity standard deviations include both within-cavity and training-seed variations and correspond to the vertical error bars in \autoref{fig4}b. These two contributions are decomposed and reported separately for each cavity in \autoref{tab:depth_statistics}. To quantify the agreement between reconstructed and reference cavity depth, we calculate the residuals by subtracting each individual prediction from the corresponding reference depth. Using tilted SEM as reference, we find a bias (mean residual) of 96 nm and an overall standard deviation of 90 nm, yielding an RMSE of 132 nm. Using AFM as reference, these values become -84 nm, 99 nm and 130 nm respectively.

Because the experimental measurements originate from only four distinct cavities, and the predictions additionally depend on the neural-network training seed, we further decompose the total residual variance into three contributions: between-cavity, within-cavity and training-seed variation. From this decomposition we obtain a training-seed standard-deviation of 63 nm, a within-cavity component of 50 nm and a between-cavity component of either 40 or 57 nm using tilted-SEM or AFM as a reference respectively. Together these values account for the overall residual variance (see section 4 of the Supporting Information for a derivation of the statistical framework). The variance decomposition shows that the training-seed component is the largest individual contributor, indicating that the reconstructed depths are sensitive to variations in the training procedure. The within-cavity spread of 50 nm describes differences in reconstructed depth between excitation positions after averaging over the training seeds. Both the training-seed and the within-cavity contributions are independent of the method used for reference depth estimation. Together these contributions highlight the potential of AR-CL to extract consistent depth information but their order of magnitude also show the need to improve robustness of the reconstruction.

Contrary to training-seed and within-cavity spread, the mean residual and between-cavity variation depend on the reference depth estimations. The uncertainties of the tilt-SEM references were determined as the half-range from manual min/max readings. Since their values (57-113 nm) are comparable to the mean residual of 96 nm, this reconstruction bias cannot be distinguished from a systematic error in the reference. The AFM derived depths have smaller estimated uncertainties (2-10 nm), but the mean values are systematically smaller than those obtained from tilted SEM, with differences ranging between 37 and 281 nm. This discrepancy may arise from limitations in accessing the cavity floor with the AFM tip due to the high aspect ratio, as well as from uncertainties in identifying the projected sidewall edges in tilted-SEM images. Such systematic effects are not necessarily captured by the reported uncertainties of either reference technique. Because the reconstructed depths lie, on average, between the two reference estimates, the RMSE values for both methods are nearly identical. However, this similarity does not establish absolute reconstruction accuracy, as the RMSE characterizes the agreement with the chosen reference rather than the true cavity depth.

\begin{table}[H]
    \centering
    \caption{Depth reconstruction results for the four investigated cavities. The upper part compares reconstructed depths with tilted-SEM and AFM reference measurements within distinct cavities. The lower part
    summarizes the pooled residual statistics and variance decomposition across all predictions.}
    \label{tab:depth_statistics}

    % Individual cavities
    \begin{tabular*}{\linewidth}{@{\extracolsep{\fill}}lcccccc@{}}
        \toprule
        Cavity &
        \shortstack{Tilt-SEM\\depth ($\upmu$m)} &
        \shortstack{AFM\\depth ($\upmu$m)} &
        \shortstack{Mean predicted\\depth ($\upmu$m)} &
        \shortstack{Per-cavity\\$\sigma_{\mathrm{total},i}$ (nm)} &
        \shortstack{Per-cavity\\$\sigma_{\mathrm{within},i}$ (nm)} &
        \shortstack{Per-cavity\\$\sigma_{\mathrm{seed},i}$ (nm)}\\
        \midrule
        I   & $0.924 \pm 0.057$ & $0.887 \pm 0.002$ & 0.879 & 81.3 & 41.4 & 70.0 \\
        II  & $1.465 \pm 0.078$ & $1.250 \pm 0.002$ & 1.336 & 73.1 & 42.0 & 59.9 \\
        III & $2.007 \pm 0.104$ & $1.818 \pm 0.006$ & 1.937 & 76.5 & 53.2 & 54.9 \\
        IV  & $2.484 \pm 0.113$ & $2.203 \pm 0.010$ & 2.343 & 90.7 & 62.0 & 66.1 \\
        \bottomrule
    \end{tabular*}

    % Pooled statistics
    \begin{tabular*}{\linewidth}{@{\extracolsep{\fill}}lcccccc@{}}
        \toprule
        Reference & $b$ (nm) & $\sigma_{\mathrm{total}}$ (nm) & RMSE (nm) & $\sigma_{\mathrm{between}}$ (nm) & $\sigma_{\mathrm{within}}$ (nm) & $\sigma_{\mathrm{seed}}$ (nm) \\
        \midrule
        Tilt-SEM & 96 & 90 & 132 & 40 & 50 & 63 \\
        AFM      & -84 & 99 & 130 & 57 & 50 & 63 \\
        \bottomrule
    \end{tabular*}

\end{table}

In addition to the uncertainty in the reference measurements, the larger residual spread observed experimentally (90-99 nm) compared to the simulated test data (37 nm) can be expected because the experimental interferograms contain effects not fully captured by the idealized FDTD library. These include FIB-induced sidewall taper and roughness, e-beam positioning uncertainty, finite signal-to-noise, parabolic mirror alignment, and collection non-uniformity. The relative contributions have not been quantified in this work and could be of interest in future studies that extend the inferred geometry. 

To place our work in the context of established dimensional metrology methods, we give an overview of the methods mentioned in the introduction in \autoref{tab:discussion_benchmarks}, showing their capabilities, performance and limitations within their main use cases. The established methods offer a higher accuracy within their respective application and the RMSE of $\sim130$ nm, relative to either reference, of this work demonstrates the need for further improvement before it could be considered for applications requiring high dimensional accuracy. Nevertheless, our results demonstrate a first step towards retrieving quantitative depth information with AR-CL, complementary to SEM imaging. The use of optical information eliminates the need for beam or stage tilt that is traditionally used in SEM and TEM 3D imaging. It offers a non-contact solution, contrary to AFM where the probe geometry could form a constraint. While confocal profilometry can reach high axial resolutions, it is diffraction limited in the lateral dimensions, whereas AR-CL measures depth complementary to the high lateral resolution of the SEM. The method presented here shares its model-based character with scatterometry, including the need for prior geometrical assumptions, and potential complications due to the non-uniqueness of the inverse problem. Within this context, the presented method offers a route towards dimensional characterization that combines both optical and electron-beam technology, motivating further investigation of its robustness and applicability to other structures and materials.

\begin{table}[H]
    \centering
    \caption{Overview of dimensional characterization methods to
    contextualize the AR-CL method presented in this work.
    Performance metrics describe particular results relevant to
    the intended application of the method, and serve as an
    indication of its capability rather than a quantitative comparison.}
    \label{tab:discussion_benchmarks}

    \small
    \setlength{\tabcolsep}{4pt}
    \renewcommand{\arraystretch}{1.12}

    % The three X-column weights must sum to 3.
    \begin{tabularx}{\linewidth}{
        @{} P{20mm} Y{0.9} Y{1.2} Y{0.9} @{}
    }
        \toprule
        \textbf{Method} &
        \textbf{Main capability} &
        \textbf{Representative performance} &
        \textbf{Main limitations} \\
        \midrule

        (CD-)SEM &
        Local dimensional measurements via a topographic contrast
        map. Height reconstruction using stage or beam tilt. &
        0.1 nm resolution, beam-tilt reaches a precision
        ($3\sigma$) of 0.81 nm on gate height and 1.49 nm on fin
        height in a FinFET metrology study
        \cite{zhang_finfet_tilt}. &
        Beam placement accuracy, contamination, beam damage,
        charging \cite{orji_roadmap}. \\
        \addlinespace[6pt]

        TEM (tomography) &
        Atomic structure and nanoscale 3D characterization. &
        0.05 nm lateral resolution \cite{orji_roadmap} and
        $\sim$1 nm 3D resolution in a tomographic tilt series
        \cite{midgley2009}. &
        Sample preparation and thickness, tilt-range restrictions,
        beam damage \cite{midgley2009, degoede_medsem}. \\
        \addlinespace[6pt]

        AFM &
        Surface topography dimensional measurements through local
        tip-sample interactions. &
        Full 3D measurement with $<1$ nm lateral,
        $<0.01$ nm vertical resolution \cite{orji_roadmap, franz_afm}. &
        Scanning speed, tip geometry can be convoluted with the
        measurement \cite{hussain_AFM}. \\
        \addlinespace[6pt]

        Confocal profilometry &
        Non-contact surface topography, height is determined from
        the axial position of peak confocal intensity. &
        Lateral resolution of $\sim$150 nm and axial resolution
        as low as 10 nm with high-NA and short-wavelength
        illumination \cite{kim2021_confocal}. &
        SNR, surface-dependent optical response, positioning and
        aberration control \cite{kim2021_confocal}. \\
        \addlinespace[6pt]

        Optical scatterometry &
        Model-based sub-wavelength optical metrology of periodic
        structures. &
        Model-dependent parameter inference of $\sim$1 nm,
        both vertical and lateral \cite{orji_roadmap}. &
        Non-uniqueness of inverse scattering problem, complicated
        modeling, optical material properties
        \cite{bodermann_scatterometry}. \\
        \addlinespace[6pt]

        AR-CL, this work &
        Model-based depth reconstruction through electron-induced optical
        information, complementary to SEM imaging. &
        Depth reconstruction demonstrated as proof-of-concept,
        with an RMSE of $\sim$130 nm, depending on the reference
        method. &
        Limited to prescribed geometries, accuracy of the forward
        model, scalability. \\

        \bottomrule
    \end{tabularx}
\end{table}

\section{Conclusion and outlook}
AR-CL spectroscopy, in combination with a library-based neural network analysis, provides detailed information on the depth of square cavities in a monocrystalline gold surface. The broad angular range of transition radiation, either direct or scattered from the sidewalls, creates interferograms with fine angular features that all represent elements of the different scattering and interference effects that occur as a result of the electron excitation. Using a neural network analysis on a library of 512 simulated interferograms we demonstrate the inverse reconstruction of cavity depth from measured AR-CL data within a prescribed cavity geometry, providing depth information complementary to traditional SEM imaging without requiring sample tilting. 

Decomposing the residual variance into training-seed, within-cavity and between-cavity contributions shows that most of the observed variation arises from the ResNet-18 training procedure (63 nm), followed by different excitation positions within the same cavity (50 nm). Both contributions are independent of the choice of reference depth. The between-cavity contribution and mean residual, however, depend on these reference measurements. Independent AFM measurements yield systematically smaller depths than tilted-SEM estimates and the RMSE of $\sim130$ nm relative to either reference therefore characterizes the agreement with the reference measurement rather than establishing absolute accuracy of the method. Variations in light collection by the parabolic mirror, e-beam positioning, and imperfections in the forward model such as a sidewall tilt or surface roughness may contribute to the observed discrepancies. Further validation on additional cavities, together with better characterization of systematic uncertainties in the reference measurements, would help to assess the quantitative performance beyond the proof of concept demonstrated here.

The current forward model describes ideal square cavities with vertical sidewalls, varying only the cavity depth and excitation position. The results demonstrate a proof of concept for depth reconstruction within the prescribed geometry but it is not established whether its variance can be distinguished from variations in other geometrical parameters. Future work could focus on improving the forward model and investigating whether additional parameters can be reconstructed independently. Furthermore, expanding the simulated training library or using a different neural network architecture may improve the robustness of the reconstruction and reduce its sensitivity to the training procedure. Polarization and hyperspectral CL maps, as well as excitation positions off the cavity symmetry axis, could also provide additional information to constrain the inverse problem.

The results demonstrate the feasibility of AR-CL-based depth reconstruction for micrometer-sized metallic cavity structures, where transition radiation and sidewall reflections generate far-field interference in the visible spectrum. An advantage of using e-beam induced radiation is that the light is created at a highly localized spot, which motivates investigation of the scalability of the method presented here to the reconstruction of sub-wavelength dimensions. Although the wavelength does not impose a direct lower limit on the precision of parameter inference, the scattering and interference physics may change substantially and may not produce a distinguishable optical fingerprint when cavity depth is varied. The achievable depth resolution would then be limited by the strength and signal-to-noise ratio of any depth-dependent change in the AR pattern, and the ability to capture this in a proper forward model. 

Similar considerations apply to an extension to dielectric or semiconductor materials, which could be an exciting direction for future work. For these materials the dominant CL generation, scattering, and contrast mechanisms may differ substantially from the Au cavity system studied here. In such systems, coherent transition radiation may be complemented by spectral information from bandgap or defect-related CL, provided that these emission mechanisms and their dependence on geometry can be accurately modelled.

\section{Methods}

\textbf{\indent Fabrication of square cavities.} The cavities were produced in a sample of bulk monocrystalline gold using focused ion beam milling with 30 keV Ga ions at a current of 0.26 nA and a pixel dwell time of 1 $\upmu$s. The shallowest cavity with a depth of $\sim0.9$ $\upmu$m was made in 300 passes and the number of passes was increased by 150 to create deeper cavities.

\textbf{AR-CL measurements.} CL experiments were performed using a ThermoFisher Apreo 2 SEM equipped with a Schottky field emission electron source operated at 30 keV and an e-beam current of 0.8 nA. CL is collected by an aluminum paraboloid mirror (acceptance angle 1.47$\pi$ sr) positioned between the sample and the electron column. By aligning the focal point of the parabolic mirror with the gold surface and the e-beam, angular information can be obtained. The CL signal is collected over an integration period of 90 s and analyzed using a Delmic SPARC system containing a Newton 920 CCD camera. The wavelength is selected with an optical band-pass filter ($\lambda = 550 \pm 25$ nm).

\textbf{Numerical simulations.} FDTD simulations were performed using Tidy3D v2.9.3 (FlexCompute Inc). Optical constants for single-crystal gold were obtained from Olmon et al. \cite{olmon2012} and the boundary of the simulation domain was defined by eight perfectly matched layers. Transition radiation was simulated by positioning a single $z$-polarized dipole source 5 nm above the vacuum-gold interface, emitting in the range between $\lambda=300$ and $\lambda=900$ nm (see section 5 of the Supporting Information for more details on this setup). To limit the simulation cost, the dipole was placed on the central $y$-axis of the cavity and PMC symmetry was applied in the plane at $x=0$. Data was recorded using a field-projection $k$-space monitor positioned several tens of nanometers above the cavity, projecting the intensity distribution onto a 101$\times$101 grid of pixels at a 500 $\upmu$m distance from the domain to the far field. The simulation runtime was chosen to be 100 fs based on a convergence test, comparing the relative simulation cost for subsequent images to their root mean square difference (RMSE) and correlation (see section 6 of the Supporting Information). Finally, the results were averaged over the range $\lambda=550 \pm 25$ nm to account for the band-pass filter used in the experiments.

\textbf{Depth inference with the neural network.} To infer the cavity depth from its AR-CL image a ResNet-18 neural network was trained using the PyTorch v2.9.0 framework installed on Python v3.12.12. Data was preprocessed by normalizing the intensity distribution of each image and splitting the 512 images in the simulation library into 70\% training, 15\% validation and 15\% test data. The training was done on batches of 10 images, resulting in 35 training steps per epoch for a total of 50 epochs. A smooth L1 loss function and an AdamW optimizer with a learning rate of $10^{-3}$ were used to update the network weights during the training loop. Validation data was used to monitor overfitting. The training procedure was repeated for 20 different random seeds used to initialize the network.

\textbf{AFM measurements.} A reference measurement of cavity depth was performed using a Bruker Dimension Icon AFM operated in PeakForce Tapping mode. Scan areas of approximately $1\times3$ $\upmu$m were selected around the cavity edge such that sufficient data on both the upper surface and the cavity floor was available. The fast-scan direction was oriented $45^\circ$ relative to the sidewall and a scan rate of $\sim0.2$ lines/s was used.

\section*{Notes}
The authors declare the following competing financial interest: Albert Polman is cofounder and co-owner of Delmic BV, a company that produces commercial cathodoluminescence systems like the one that was used in this work.

\section*{Acknowledgements}
The authors thank Jian Gao for assistance with AFM measurements, and Wiebke Albrecht, Ethan Kensett, Timon Fliervoet and Hang-Kwang Nienhuys for insightful discussions. This work is part of the research program of the Dutch Research Council (NWO) and was performed at the research institutes AMOLF and ARCNL. The Advanced Research Center for Nanolithography ARCNL is a public–private partnership between the University of Amsterdam, Vrije Universiteit Amsterdam, Rijksuniversiteit Groningen (RUG), NWO, and the semiconductor-equipment manufacturer ASML. This work has received funding from the European Research Council (ERC) under Grant Agreement No. 101019932 (QEWS), the European Innovation Council (EIC) under Grant Agreement No. 101017720 (EBEAM).

%%%%%%%%%%%%%%%%%%%%%%%%%%%%%%%%%%%%%%%%%%%%%%%%%%%%%%%%%%%%%%%%%%%%%
%% If you are using classical BibTeX rather than biblatex,
%% remove the \printbibliography and uncomment the \bibliograpy one
%%%%%%%%%%%%%%%%%%%%%%%%%%%%%%%%%%%%%%%%%%%%%%%%%%%%%%%%%%%%%%%%%%%%%
\printbibliography
%\bibliography{acs-template.bib}

\end{document}

\maketitle


\section{Fourier peak analysis of an AR-CL image}
\begin{figure}[H]
    \centering
    \includegraphics[width=\linewidth]{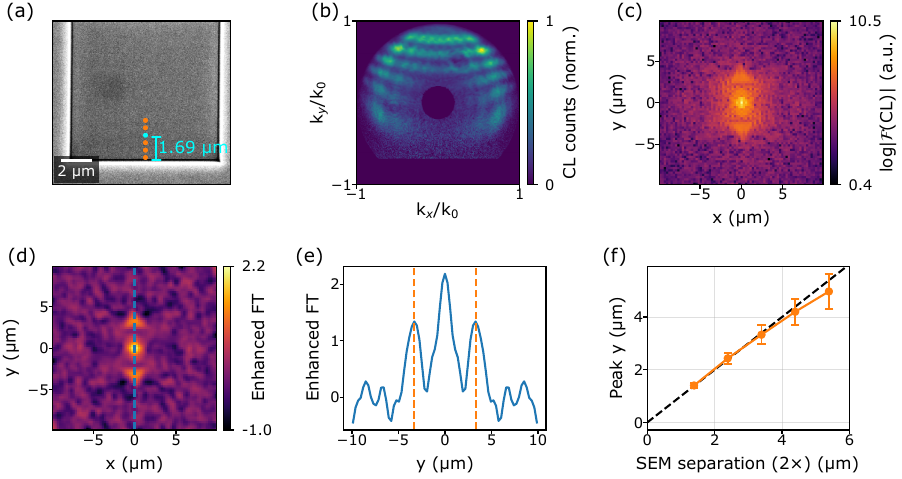}
    \caption{(a) Zoomed-in SEM image of a square cavity. The dots indicate the locations where AR-CL data was taken, the first location being $\sim$200 nm from the lower sidewall and each subsequent dot spaced 500 nm further inward. Data for the cyan dot are shown in figures (b-e). (b) AR-CL interference image taken at the location of the cyan dot in the panel (a). (c) Fourier transform of the AR-CL pattern. (d) Enhanced image of the Fourier transform from (c) using Gaussian filtering. The visibility of the additional peaks is enhanced, making it easier for an algorithm to identify its coordinates. (e). Linecut through (d) at x=0. The orange line indicates the peak centroid at $y\approx \pm3.4$ $\upmu$m. (f) Graph showing the correlation between the interference source separation inferred from the Fourier transform and a characteristic dimension measured in the SEM image. The SEM data shows $2\times$ the distance between the excitation position and the lower sidewall. The error bars represent the peak width.}
    \label{fig:supp_fourier}
\end{figure}

\autoref{fig:supp_fourier} shows the analysis flow used to determine the origin of the interference fringes. In the SEM picture shown in \autoref{fig:supp_fourier}a, the distance between the cathodoluminescence (CL) excitation and the lower sidewall of the cavity is determined. For each excitation point, the angle-resolved (AR) CL interferogram is recorded and its Fourier transform is calculated. \autoref{fig:supp_fourier}a and \autoref{fig:supp_fourier}c show these data respectively for the excitation position marked in \autoref{fig:supp_fourier}a. The coordinates of the second-order maximum in this Fourier image correspond to the separation between the two interfering light sources. To accurately track the location of this maximum, the image is enhanced by subtracting a broadly Gaussian-smoothed ($\sigma=5$) version of the image from a narrow ($\sigma=1$) Gaussian-filtered version as shown in \autoref{fig:supp_fourier}d. The $y$-coordinates of the peaks are then determined from a linecut at $x=0$, shown in \autoref{fig:supp_fourier}e using the \texttt{scipy.signal.find\_peaks} function. The width of the peaks is then determined using \texttt{scipy.signal.peak\_widths} at half prominence. Because the peaks can be broadened and asymmetric, rather than using the $y$-coordinate of their maximum, we determine the amplitude-weighted centroid, given by
\begin{equation}
    \bar y=\frac{\sum^N_iw_iy_i}{\sum^N_iw_i},
\end{equation}
where $w_i$ is the background-subtracted amplitude of the linecut, and the summation runs over $N$ datapoints within the width of the peak. The Fourier-derived separation is then obtained by averaging the centroid positions of the positive and negative peaks.

For each location where AR-CL data was recorded, this characteristic distance obtained from Fourier analysis is compared to the dimensions known from the SEM image, which is plotted in \autoref{fig:supp_fourier}f. The error bars are determined from the Fourier peak width according to
\begin{equation}
    \sigma_\mathrm{FFT}=\sqrt{\frac{\sum_i^Nw_i(y_i-\bar y)^2}{\sum_i^Nw_i}}.
\end{equation}
The data for the point closest to the sidewall is left out as the second Fourier peak could not be distinguished from the central maximum. A correlation is found when multiplying the distance between the CL excitation point and the lower sidewall by a factor 2, with the first three datapoints showing a linear behavior which starts to deviate as the beam is moved further from the sidewall. 

\begin{figure}[H]
    \centering
    \includegraphics[width=\linewidth]{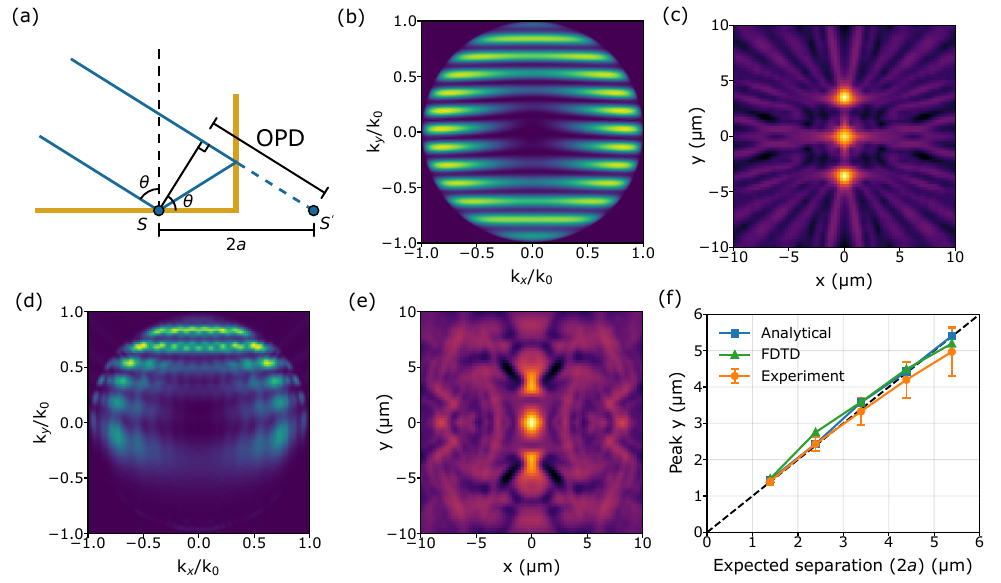}
    \caption{(a) Diagram illustrating the interference between a light source $S$ and its virtual mirror source $S'$. The optical path difference between the two rays recorded at angle $\theta$ in the far-field is given by $\mathrm{OPD}=2a\sin\theta$. (b) Analytically calculated interferogram for two transition radiation sources spaced a distance $2a$ with $a=1.69$ from each other. The fringe count is similar to what is observed in the experiment but fringe curvature, shadowing effects and subtle features aren't caught by this model. (c) Gaussian enhanced Fourier transform of (b). (d) FDTD interferogram simulated by placing a $z$-oriented dipole 1.69 $\upmu$m from the lower sidewall in a $10\times 10\times 2.4$ $\upmu\mathrm{m}$ Au cavity. The full 3D simulation more accurately reconstructs the subtle features observed in the experiment. (e) Gaussian enhanced Fourier transform of (d). All images (a-d) are normalized between 0 and 1. (f) Centroid $y-$position of the second Fourier peak plotted against $2\times$ the distance between the excitation point and the cavity sidewall. Data for $a=0.19$ is left out as the second Fourier peak could not be distinguished from central peak.}
    \label{fig:supp_interference_diagram}
\end{figure}

The linear trend can be explained by identifying reflection of the primary transition radiation from the sidewall as the source of the interference. A geometrical construction, shown in \autoref{fig:supp_interference_diagram}, shows that the optical path difference can be written as
\begin{equation}
    \mathrm{OPD} = 2a\sin\theta,
\end{equation}
with $a$ the distance between the electron beam and the cavity sidewall. The phase difference between the two rays is $\Delta\phi=\frac{2\pi}{\lambda}\mathrm{OPD}$ and using the fact that in the normalized coordinate system of the AR interferograms $k_y=\sin\theta$, we find that the phase difference can be written as
\begin{equation}
    \Delta\phi = \frac{2\pi}{\lambda}2ak_y.
\end{equation}
Furthermore, the intensity of the interferogram can be written as
\begin{equation}
    I_\mathrm{total} = I_{\mathrm{TR}} + I_{\mathrm{refl}}+2\sqrt{ I_{\mathrm{TR}}I_{\mathrm{refl}}}\cos(\Delta\phi),
\end{equation}
showing that a perfect reflection would yield an interferogram with fringes with a frequency $\frac{2a}{\lambda}$. A Fourier transform of the interferogram, scaled by the source wavelength, would thus show a peak at $y=2a$.

In Ref.\cite{brenny2016} it is shown that the electric field of the transition radiation generated by an electron with velocity $v$ impinging on a planar interface between two media with dielectric functions $\epsilon_1$ and $\epsilon_2$ respectively is given by
\begin{equation}
\label{TR_fields}
    \mathbf{E}_\mathrm{TR}(r, \theta, \omega)=ik\cos\theta D\mu_1\frac{e^{ikr}}{r}\hat\theta,
\end{equation}
with 
\begin{equation}
    D=\frac{2ieQ/c}{\epsilon_1q_{z2}+\epsilon_2q_{z1}}
\end{equation}
and
\begin{equation}
    \mu_1 = \frac{\epsilon_1q_{z2}-\epsilon_1\omega/v}{q^2-k^2\epsilon_2} - \frac{\epsilon_1q_{z2}-\epsilon_2\omega/v}{q^2-k^2\epsilon_1},
\end{equation}
where $Q$ is the momentum parallel to the interface and $q_{z1}$ and $q_{z2}$ the axial momentum in the respective media. Using \eqref{TR_fields} we can analytically calculate the expected far-field interference pattern for transition radiation reflected in an infinite planar mirror by placing two sources of the electric field at a distance $2a$. The coherent sum $I(\lambda)=|E_\mathrm{TR}(\lambda)+r_{\mathrm{fresnel}}E_{refl}(\lambda)|^2$ is simulated for different wavelengths within the band $\lambda\in[525, 575]$ with $\delta\lambda=2.5$ nm. The resulting interferogram is then given by the incoherent sum over the bandwidth. \autoref{fig:supp_interference_diagram}b shows the result for $a=1.69$ $\upmu$m. For $k_y>0$ this interferogram shows the same fringe count as the experimental interferogram from \autoref{fig:supp_fourier}b, further strengthening the presumption that reflection off the cavity sidewall causes the interference. \autoref{fig:supp_interference_diagram}c shows the Gaussian enhanced Fourier transform of the analytical interferogram. We repeat this calculation for each distance $a$ used in the experiment and, following the same procedure as previously, determine the centroid of the Fourier-peak. We show the $y-$coordinate of the different centroids in the blue curve in \autoref{fig:supp_interference_diagram}(f) and find that, while it closely follows its programmed curve of $y=2a$, some deviations are still present. This shows that even in an analytical calculation the extracted centroid can differ when the Fourier features are broadened or asymmetric.

This analytical calculation however was done by simply placing two light sources at a distance $2a$ from each other. This assumes that the lower cavity sidewall is an infinite planar mirror, thereby not capturing any shadowing effects, edge scattering or reflections of the other sidewalls caused by the full 3D geometry of the cavity. These effects likely contribute significantly to the experimental interferogram, motivating the use of a full FDTD simulation. In \autoref{fig:supp_interference_diagram}d, we show the FDTD result for a dipole placed $1.69$ $\upmu$m above the cavity sidewall. We note that for $k_y>0$ the amount of fringes in the experimental, analytical and FDTD simulation is similar, but the FDTD simulation more closely resembles changes in fringe thickness and curvature seen in \autoref{fig:supp_fourier}b. The enhanced Fourier transform of (d) is shown in \autoref{fig:supp_interference_diagram}e and shows, apart from the second order peak that we have been tracking before, many other features likely relating to the finer fringes present in the interferogram. These features are however difficult to see in the experimental data as they are likely obscured by the background noise. 

\autoref{fig:supp_interference_diagram}(f) shows the Fourier-derived separations obtained from the analytical, FDTD and experimental interferograms. For clarity, the peak widths are only indicated for the experimental data, but comparable widths are observed in the simulated datasets. All three curves approximately follow the linear trend, although small deviations from ideal $(2a)$ remain. This motivates reflection of transition radiation from the cavity sidewall as a likely explanation for the cause of the interference. However, the Fourier peak widths and asymmetries make it difficult to interpret their centroid position as exact source separations. Furthermore, while the simplified mirror model explains the fringe density, it does not capture additional features introduced by the finite cavity geometry, such as fringe curvature and shadowing. In the main text it is shown that these features correlate with cavity depth, but they are difficult to relate directly to an individual Fourier component. Therefore, a data-driven approach using a convolutional neural network, known for its strength in image processing, was chosen to extract cavity depth information from angle-resolved CL measurements.

\section{ResNet-18 learning curve}
\begin{figure}[H]
    \centering
    \includegraphics[width=0.5\linewidth]{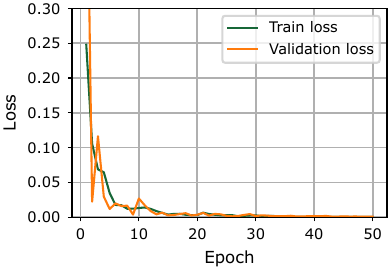}
    \caption{Training and validation loss of a representative ResNet-18 neural network, initialized with random seed 19, over the course of 50 epochs.}
    \label{fig:loss}
\end{figure}

\autoref{fig:loss} shows the learning curve of a single ResNet-18 neural network over the course of 50 epochs. The training loss decreases to a value $<10^{-3}$ within 33 epochs. The loss on the validation dataset was tracked to prevent overfitting of the network. The epoch with the lowest validation loss was used to perform inference on the simulated test data and on the experimental data.

\section{Reference depth characterization}
\label{sec:SEM_tilt}
\begin{figure}[H]
    \centering
    \includegraphics[width=0.5\linewidth]{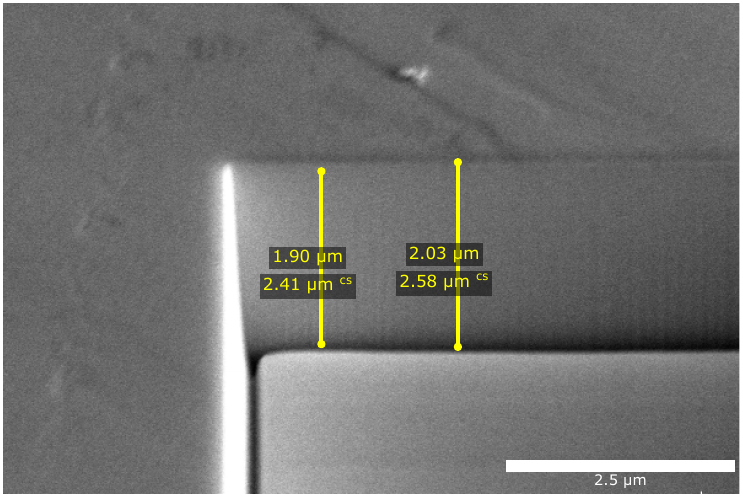}
    \caption{Cavity depth measurement via sidewall projection under a 52$^\circ$ tilt angle. The two line measurements show the minimum and maximum estimate while the labeled numbers show the projected depth and the tilt-angle corrected depth.}
    \label{fig:SEreference}
\end{figure}

To establish reference depth estimates for evaluating the library-based AR-CL reconstruction method, we characterized the cavities using both tilted-SEM and atomic force microscopy (AFM). In tilted-SEM, the depths are estimated from secondary electron images recorded while viewing the cavity under a 52$^\circ$ tilt angle. The depth is then related to its projection via
\begin{equation}
    d = L_\mathrm{proj} / \sin(52^\circ).
\end{equation}
For each cavity we estimate a minimum and a maximum depth from the SE image, one example of which is shown in \autoref{fig:SEreference}. These measurements were performed on the cavity sidewall closest to the CL excitation position. In order to correctly estimate the top of the sidewall the measurements needed to be performed in one of the corners of the cavity. Two pairs of minimum and maximum estimates were obtained at each of the two corners. The reference values reported in Figure 4 of the main text are the center of these measurements, given by $d_\mathrm{mean}=\frac{\mathrm{mean}(d_\mathrm{max}) + \mathrm{mean}(d_\mathrm{min})}{2}$ , while the reported error is $d_\mathrm{error}=\frac{\mathrm{mean}(d_\mathrm{max}) -\mathrm{mean}(d_\mathrm{min})}{2}$. Values for individual cavity measurements and their center and error are reported in \autoref{tbl:measurements}. The reported error represents the half-range between the manually estimated minimum and maximum depths and should therefore be interpreted as an estimate of the uncertainty associated with identifying the sidewall boundaries, rather than a statistical standard deviation or confidence interval.

\begin{table}[H]
  \centering
  \caption{Tilt-angle corrected depth measurements of the four cavities studied in these experiments. Two min/max estimates were taken in both the left and right corner of the sidewall. For each cavity the center and error of the min/max range are reported.}
  \label{tbl:measurements}
  \centering
  \setlength{\tabcolsep}{4pt}
  \begin{tabular}{|l|llll|llll|llll|llll|}

     \hline
     Cavity & \multicolumn{4}{|c|}{I} & \multicolumn{4}{c|}{II} & \multicolumn{4}{c|}{III} & \multicolumn{4}{c|}{IV} \\
     \hline
      Corner & \multicolumn{2}{|c}{L} & \multicolumn{2}{|c|}{R} & \multicolumn{2}{|c|}{L} & \multicolumn{2}{c|}{R} & \multicolumn{2}{|c|}{L} & \multicolumn{2}{c|}{R} & \multicolumn{2}{|c|}{L} & \multicolumn{2}{c|}{R} \\
     \hline
     Min ($\upmu$m)  & 0.87 & 0.89 & 0.85 & 0.87 & 1.35 & 1.42 & 1.38 & 1.40 & 1.92 & 1.87 & 1.88 & 1.94 & 2.35 & 2.35 & 2.37 & 2.41\\
     Max ($\upmu$m) & 0.99 & 0.95 & 0.97 & 1.02 & 1.49 & 1.59 & 1.50 & 1.59 & 2.12 & 2.05 & 2.12 & 2.15 & 2.57 & 2.59 & 2.65 & 2.58\\
     Center ($\upmu$m) & \multicolumn{4}{|c|}{0.92 $\pm$ 0.06} & \multicolumn{4}{c|}{1.47 $\pm$ 0.08} & \multicolumn{4}{c|}{2.01$ \pm$ 0.10} & \multicolumn{4}{c|}{2.48 $\pm$ 0.11} \\

     \hline
    \end{tabular}
\end{table}

\begin{figure}[H]
    \centering
    \includegraphics[width=\linewidth]{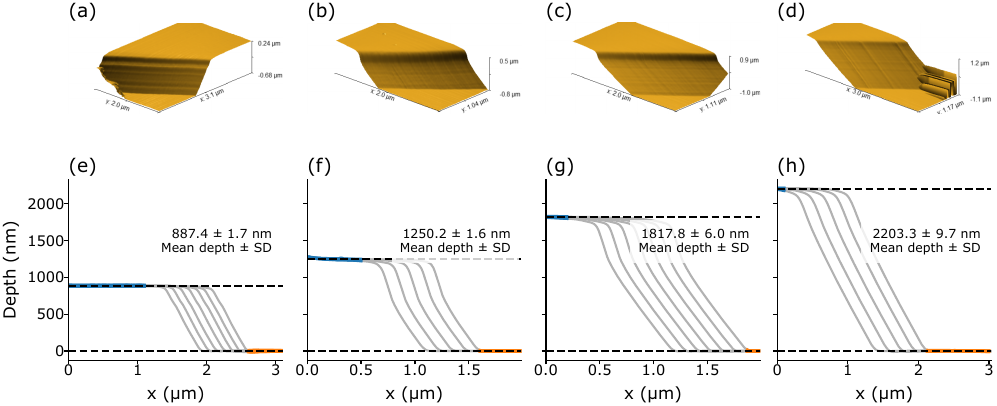}
    \caption{AFM characterization of cavity depth. (a-d) AFM topography maps recorded over a $\sim 1\times 3$ $\upmu$m area across the cavity sidewall. The broad apparent sidewall and irregularities at the bottom are attributed to measurement artifacts. (e-h) Corresponding height profiles, averaged over 10 scan lines, extracted such that both the upper surface and floor are included. Blue and orange segments indicate the region used to fit the plateau regions, indicated by the dashed lines, from which the depth is extracted.}
    \label{fig:afm_data}
\end{figure}

To independently characterize the cavity depths and further assess the SEM-derived estimates, additional AFM measurements were performed on the sidewall. Because the cavities sidewalls are relatively steep and deep for an AFM we use a low scan rate of 0.2 lines/s and limit the scan area around the edge to $\sim1\times 3$ $\upmu$m. To facilitate tracking the steep height transition the fast-scan direction is oriented $\sim 45^\circ$ relative to the sidewall. The resulting topography maps are shown in \autoref{fig:afm_data}(a-d). A broad height transition, extending over more than 1 $\upmu$m, is observed,  which is not observed in the SEM images of the same cavities. We therefore attribute this broadening to AFM measurement artifacts, arising from the finite tip geometry and  limited feedback response when tracking the steep transition. Additionally, the maps in panels (a) \& (d) show irregularities in the initial scan lines, suggesting feedback instabilities when tracking the steep sidewalls. Nevertheless, the height profiles eventually reach a relatively flat region which we identify as the apparent cavity floor.

After plane leveling using the surrounding gold surface, several $(x,z)$ profiles were extracted from each topography map, each averaged over 10 adjacent scan lines along the $y$-direction. The profiles were selected such that both the upper gold surface and the cavity floor were included, as shown in \autoref{fig:afm_data}(e-h). The depth is then extracted by fitting a horizontal line to the plateau regions overlapping in each profile and taking the difference. We identify depths of $0.887, 1.250, 1.818$ and $2.203$ $\upmu$m, with standard deviations of $1.7, 1.6, 6.0$ and $9.7$ nm respectively. These standard deviations describe variation between the extracted profiles rather than absolute measurements of uncertainty.

The AFM-derived depths are systematically smaller than the tilted-SEM estimates, with differences between $37$ and $281$ nm. Although the AFM profiles show relatively small variations in the measured step heights, systematic uncertainties associated with the high-aspect ratio of the edge measurement cannot be excluded. In particular, the finite tip geometry and feedback response may affect the measurement of the cavity floor. Similarly, the tilted-SEM measurements rely on manually identifying the sidewall boundaries, which introduces uncertainty in the measured sidewall projection. The origin of the discrepancy is not conclusively established and, consequently, both measurements should be treated as reference estimates rather than independently verified ground-truth depths. 

\section{Statistical framework of experimental depth reconstruction}
To quantify the agreement between reconstructed and reference depth, we perform a residual analysis of the experimental predictions. Because the measurements within the same cavity share the same reference depth, and each image is evaluated using multiple independently trained neural networks initialized with a different random seed, we make a decomposition of the within-cavity, the between-cavity, and the training-seed variation. 

Defining $\hat d_{ijs}$ as the prediction from interferogram $j$ in cavity $i$, obtained using neural network $s$, and $d_i^\mathrm{ref}$ as the reference estimate of cavity $i$, we write the residuals as
\begin{equation}
 \label{eq:residuals}
    r_{ijs}=d_i^\mathrm{ref} - \hat d_{ijs}.
\end{equation}
The mean residual and total variance of the pooled residuals are then given by
\begin{equation}
    b = \frac{1}{N}\sum^M_{i=1}\sum^{n_i}_{j=1}\sum_{s=1}^Sr_{ijs}
\end{equation}
and
\begin{equation}
    \label{eq:sigma_total}
    \sigma_\mathrm{total}^2 = \frac{1}{N}\sum^M_{i=1}\sum^{n_i}_{j=1}\sum_{s=1}^S(r_{ijs}-b)^2,
\end{equation}
with $S$ the number of independent networks, $M$ the number of distinct cavities, and $n_i$ the number of measurements within cavity $i$. Each image is evaluated using the same $S$ networks, giving a total number of predictions $N=S\sum^M_{i=1}n_i$.

To decompose the total variance, we first define the mean residual
for each interferogram, averaged over the training seeds,

\begin{equation}
    \bar{r}_{ij} = \frac{1}{S}
    \sum_{s=1}^{S} r_{ijs},
\end{equation}

and the mean residual within a single cavity,

\begin{equation}
    \bar{r}_i = \frac{1}{n_i}
    \sum_{j=1}^{n_i}\bar{r}_{ij}.
\end{equation}

We can then write

\begin{equation}
    r_{ijs}-b =
    (r_{ijs}-\bar{r}_{ij})
    +(\bar{r}_{ij}-\bar{r}_i)
    +(\bar{r}_i-b)
\end{equation}

and plug its square into \eqref{eq:sigma_total}. The cross terms will vanish when taking the summation, since

\begin{equation}
    \sum_{s=1}^{S}(r_{ijs}-\bar{r}_{ij})=0
    \quad\text{and}\quad
    \sum_{j=1}^{n_i}(\bar{r}_{ij}-\bar{r}_i)=0.
\end{equation}

The total variance can therefore be written as

\begin{equation}
    \sigma_{\mathrm{total}}^2 = \frac{1}{N}\sum_{i=1}^{M}\sum_{j=1}^{n_i}\sum_{s=1}^{S}(r_{ijs}-\bar{r}_{ij})^2 + \frac{S}{N}\sum_{i=1}^{M}\sum_{j=1}^{n_i}(\bar{r}_{ij}-\bar{r}_i)^2 + \frac{S}{N}\sum_{i=1}^{M}n_i(\bar{r}_i-b)^2.
\end{equation}

The first term represents the variation between predictions on the same image from independently trained neural networks. We identify this as the training-seed variance,
\begin{equation}
    \sigma_{\mathrm{seed}}^2 =
    \frac{1}{N}
    \sum_{i=1}^{M}\sum_{j=1}^{n_i}\sum_{s=1}^{S}
    (r_{ijs}-\bar{r}_{ij})^2.
\end{equation}

The second term describes the variation between different
excitation positions within the same cavity, after averaging
over the training seeds. We identify this as the pooled
within-cavity variance,

\begin{equation}
    \sigma_{\mathrm{within}}^2 =
    \frac{S}{N}
    \sum_{i=1}^{M}\sum_{j=1}^{n_i}
    (\bar{r}_{ij}-\bar{r}_i)^2.
\end{equation}

Using the definition of the residual, we find that

\begin{align}
    \bar{r}_{ij}-\bar{r}_i
    &= (d_i^{\mathrm{ref}}-\bar{\hat{d}}_{ij})
    -(d_i^{\mathrm{ref}}-\bar{\hat{d}}_i)
    \nonumber\\
    &= -(\bar{\hat{d}}_{ij}-\bar{\hat{d}}_i),
\end{align}

where $\bar{\hat{d}}_{ij}$ is the prediction for interferogram
$j$ averaged over all training seeds, and $\bar{\hat{d}}_i$ is
the mean prediction for cavity $i$.

By taking the square, we find that the within-cavity variance
can be expressed as

\begin{equation}
    \sigma_{\mathrm{within}}^2 =
    \frac{S}{N}
    \sum_{i=1}^{M}\sum_{j=1}^{n_i}
    (\bar{\hat{d}}_{ij}-\bar{\hat{d}}_i)^2.
\end{equation}

Similarly, for the training-seed contribution we find

\begin{equation}
    r_{ijs}-\bar{r}_{ij}
    =-(\hat{d}_{ijs}-\bar{\hat{d}}_{ij}).
\end{equation}

Both the within-cavity and training-seed variances are therefore independent of the reference depth estimates. Together, these two contributions describe the variation in predicted depths within an individual cavity, which can be written as
\begin{align}
    \sigma_i^2 &=
    \sigma_{\mathrm{within},i}^2 +
    \sigma_{\mathrm{seed},i}^2 \\
    &=
    \frac{1}{n_iS}
    \sum_{j=1}^{n_i}\sum_{s=1}^{S}
    (\hat{d}_{ijs}-\bar{\hat{d}}_i)^2.
\end{align}
Here, the variance components are evaluated for cavity $i$ only, rather than pooled across all cavities. The standard deviation $\sigma_i$ is used for the vertical error bars in
Figure 4(b) in the main text, and it represents the combined variation across excitation positions and independently trained networks.

Finally, we identify the third term as the between-cavity variance,

\begin{equation}
    \sigma_{\mathrm{between}}^2 =
    \frac{S}{N}
    \sum_{i=1}^{M}n_i(\bar{r}_i-b)^2.
\end{equation}

Combining these three contributions yields the final
decomposition,

\begin{equation}
    \boxed{
    \sigma_{\mathrm{total}}^2 =
    \sigma_{\mathrm{within}}^2 +
    \sigma_{\mathrm{between}}^2 +
    \sigma_{\mathrm{seed}}^2.
    }
\end{equation}

The quantitative decomposition is reported in the main text. The within-cavity variance quantifies differences between the seed-averaged predictions at different excitation positions, while the training-seed variance quantifies the variability of predictions between independently trained networks for the same interferogram. The latter includes both systematic differences between network predictions and image-dependent variations. Both contributions are independent of the chosen reference depths. In contrary, the between-cavity variance quantifies variation of the cavity-averaged residuals around the overall mean residual. Both the mean residual and the between-cavity variance depend on the reference depth and may therefore be affected by errors in the tilted-SEM min/max estimations or the AFM analysis described in \autoref{sec:SEM_tilt}.

\section{FDTD simulation setup}

\begin{figure}[H]
    \centering
    \includegraphics[width=\linewidth]{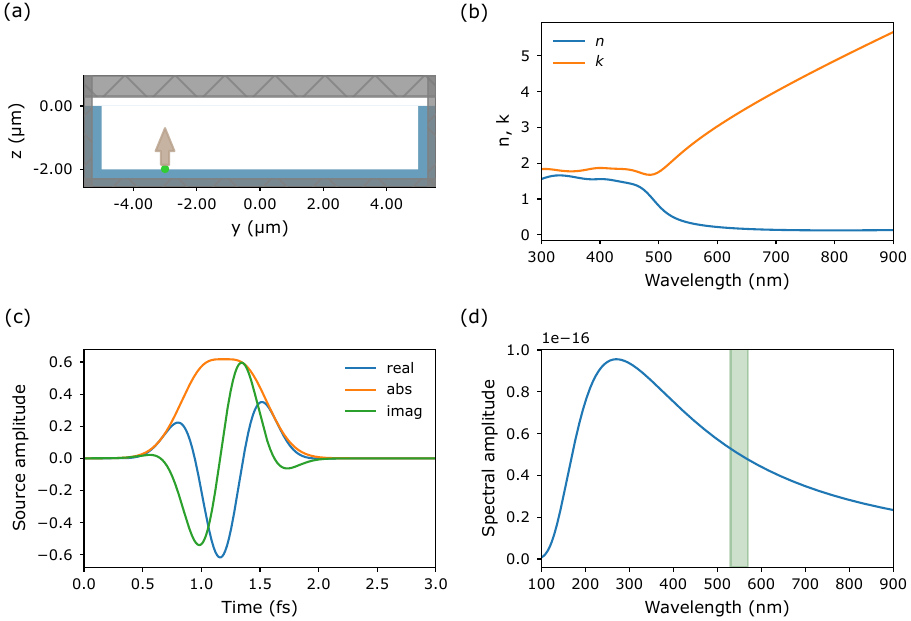}
    \caption{FDTD simulation setup. (a) A $z$-oriented dipole is placed on the $(0, y)$ axis inside a $10 \ \upmu \mathrm{m}\times10 \ \upmu \mathrm{m}\times z$ cavity. A field-projection $k$-space monitor is positioned several tens of nanometers above the cavity and the whole simulation domain is bounded by a perfectly matched boundary layer. (b) Refractive index dispersion obtained from a complex pole-residue fit to ellipsometry data taken from Ref.\cite{olmon2012} used for the monocrystalline gold material in which the cavity is milled. The cavity is defined as the vacuum permittivity ($n=1$). (c) Gaussian pulse used to excite the dipole source. (d) Radiated spectrum defined by the excitation source. The green band indicates the wavelengths of interest within the experimental optical filter of $550\pm25$ nm.}
    \label{fig:tidy3dsetup}
\end{figure}

\autoref{fig:tidy3dsetup} shows the setup of the Finite-Difference Time-Domain (FDTD) simulations in Tidy3D that were used to construct the training data library of AR-CL images. The library is constructed by varying the $y$-coordinate of the dipole and the cavity depth $z$ shown in (a). Because monocrystalline gold is a dispersive material, its permittivity needed to solve Maxwell's equations is obtained by making a pole-residue fit to ellipsometry data according to 
\begin{equation}
    \epsilon(\omega) = \epsilon_\infty - \sum_k\bigg[\frac{c_k}{i\omega+a_k}+\frac{c^*_k}{i\omega+a^*_k}\bigg],
\end{equation}
where $\epsilon_\infty=1$ for a Drude metal in vacuum, $a_k$ is the pole location and $c_k$ its residue. The resulting dispersion for ellipsometric data taken from Olmon et al. \cite{olmon2012} is shown in (b). The dipole source was excited with a Gaussian pulse in the time domain such that it emits with a broadband spectrum in the frequency domain, similar to transition radiation, as shown in \autoref{fig:tidy3dsetup}(c,d). 

\section{FDTD simulation convergence testing}

Since we are simulating the scattering of light inside a cavity, it is important that we let the FDTD simulation run long enough for the optical fields to fully decay. A longer runtime, however, leads to a higher computational cost which in the case of Tidy3D is billed using a credit system called FlexCredits\footnote{https://docs.flexcompute.com/projects/tidy3d/en/stable/faq/docs/faq/what-is-a-flexcredit.html}. For the purpose of this study we do not necessarily need all fields to fully decay. We are only interested in relevant contributions to the $k$-space monitor which are likely to affect the neural network performance. Therefore, we choose to run a convergence test of 3 simulations lasting 50, 100 and 200 fs respectively. We then compare the resulting interference images for 50 vs 100 and 100 vs 200 fs runtime. 

\begin{figure}[H]
    \centering
    \includegraphics[width=\linewidth]{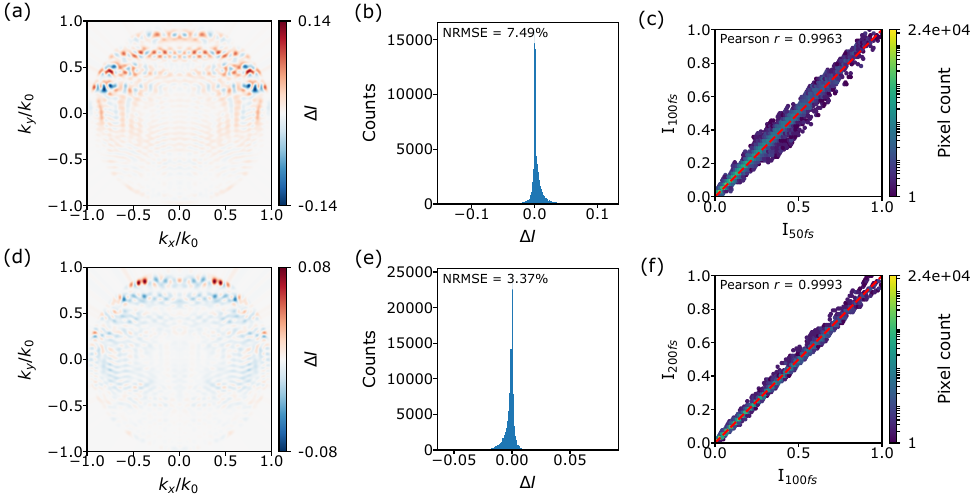}
    \caption{Comparison between FDTD simulation results for (a-c) 50 vs 100 fs runtime and (d-f) 100 vs 200 fs runtime. (a) \& (d) show the intensity difference between the subsequent images. (b) \& (e) show the histogram of these differences and reports their normalized root mean square error. (c) \& (f) show the correlation of the intensity value per pixel. The colorbar indicates the pixel density on the plot.}
    \label{fig:convergence}
\end{figure}

The results of this comparison are shown in \autoref{fig:convergence}. In \autoref{fig:convergence}a we show the intensity difference between the 100 fs and 50 fs simulations. This difference is at most 14\% and shows up mainly in the high-frequency components. In \autoref{fig:convergence}b we plot a histogram of these intensity differences and, using the Python package \texttt{skimage}, we calculate the normalized root mean square error of 7.49 \%, according to
\begin{equation}
    \mathrm{NRMSE}=\sqrt{\frac{1}{M\times N}\sum_{i=1}^M\sum_{j=1}^N(I_1(i,j)-I_2(i,j))^2} \times 100\%,
\end{equation}
with $M\times N$ the dimensions of an image. In \autoref{fig:convergence}c we show the correlation of each individual pixel and report a Pearson correlation coefficient of $r=0.9963$. This analysis is repeated for the comparison between 200 and 100 fs runtime in Figures \ref{fig:convergence}(d-f) where we find a normalized RMSE of 3.37\% and a correlation of 0.9993. 

The comparison metrics are further summarized in \autoref{tab:relative_cost}, where we inspect them against the relative simulation cost. Switching from 50 fs to 100 fs runtime costs 1.32 times as many credits while switching from 100 to 200 fs runtime costs 1.53 as many credits. For the metrics we use the rule of thumb that two images with a normalized RMSE $< 5\%$ and a correlation $>0.99$ are considered similar. For this reason we conclude that the simulations with 100 fs and 200 fs runtime give similar enough results for our purposes, while simulations running for 100 fs do yield a significant improvement over those running for 50 fs. For building the simulation library a simulation time of 100 fs was therefore used.

\begin{table}[H]
    \centering
    \caption{Relative FlexCredits costs and image comparison metrics for subsequent FDTD simulations with different runtime.}
    \begin{tabular}{|c c c c c|}
        \hline
         Runtime & FlexCredit cost & Relative cost & nRMSE & Pearson correlation  \\
         \hline
          50 fs & 0.047 & 1.00$\times$ & - & - \\ 
          100 fs & 0.062 & 1.32$\times$ & 7.49\% & 0.9963 \\ 
          200 fs & 0.095 & 1.53$\times$ & 3.37\% & 0.9993 \\ 
    \hline
    \end{tabular}
    \label{tab:relative_cost}
\end{table}

\printbibliography